\documentclass{nature}
\usepackage{graphicx}
\usepackage{amssymb}
\usepackage{float}
\usepackage{bm}
\usepackage{array}
\usepackage{amsmath}
\usepackage{booktabs} % Required for horizontal rules in tables
\usepackage{multicol}

\graphicspath{{Pic/}}

\title{Scaling Laws in Earthquake Memory for Interevent Times and Distances}

\author{Yongwen Zhang$^{1,2,3}$, Jingfang Fan$^{4,2}$, Warner Marzocchi$^{5}$, Avi Shapira,$^{6}$ \\ Rami Hofstetter,$^{7}$ Shlomo Havlin,$^{2}$ and Yosef Ashkenazy$^{1}$}

\begin{document}

\maketitle

\begin{affiliations}
 \item Department of Solar Energy and Environmental Physics, The Jacob Blaustein Institutes for Desert Research, Ben-Gurion University of the Negev, Midreshet Ben-Gurion 84990, Israel;
 \item Department of Physics, Bar-Ilan University, Ramat Gan 52900, Israel;
 \item Data Science Research Center, Faculty of Science, Kunming University of Science and Technology, Kunming 650500, Yunnan, China;
 \item Potsdam Institute for Climate Impact Research, 14412 Potsdam, Germany;
 \item Department of Earth, Environmental, and Resources Sciences, University of Naples, Federico II, Complesso di Monte Sant’Angelo, Via Cinthia, 21 80126 Napoli, Italy;
 \item National Institute for Regulation of Emergency and Disaster, College of Law and Business, Bnei Brak, 511080, Israel;
 \item Geophysical Institute of Israel, Lod 7019802, Israel.
  
 \end{affiliations}

\begin{abstract}

Over the past decades much effort has been devoted towards understanding and forecasting natural hazards. However, earthquake forecasting skill is still very limited and remains a great scientific challenge\cite{Jordan2011}. The limited earthquake predictability is partly due to the erratic nature of earthquakes and partly to the lack of understanding the underlying mechanisms of earthquakes\cite{DeArcangelis2016}. To improve our understanding and potential forecasting, here we study the spatial and temporal long-term memory of interevent earthquakes above a certain magnitude using lagged conditional probabilities.
  % Both the spatial memory and the lagged conditional probability are studied here for the first time.
We find, in real data, that the lagged conditional probabilities show long-term memory for both the interevent times and interevent distances and that the memory functions obey scaling and decay slowly with time, while, at a characteristic time, the decay crossesover to a fast decay. 
% We find that the memory of both the interevent times and distances obey scaling laws.
We also show that the ETAS model, which is often used to forecast earthquake events, yields scaling functions of the temporal and spatial interevent intervals which are not consistent with those of real data. 

\end{abstract}

Earthquakes involve complex processes that span over a wide range of spatial and temporal scales\cite{DeArcangelis2016}. There are two well known empirical laws regarding earthquakes: (i) the Gutenberg-Richter law which determines the relation between the number of earthquakes $N$ in a given region and a time period and the magnitude $m$ as, $N(m) \propto 10^{-bm}$ ($b \approx 1$)\cite{Gutenberg1944a}, (ii) the Omori law according to which the rate of triggered events is $\sim t^{-p}$, where $t$ is the time since a triggering earthquake ($p \approx 1$ for large earthquakes)\cite{Utsu2011}. 

Using the Gutenberg-Richter law and the exponent of the Omori law, Bak et al.\cite{Bak2002} found that the probability density function (PDF) of interevent times for different magnitude thresholds and different spatial grid sizes can be rescaled into a single function. This suggests a universal scaling law for earthquakes. Corral\cite{Corral2003} extended this scaling to different regions and introduced a more general approach. Specifically, he introduced a unified function, $f(x)$, to describe the distribution of interevent times as $D(\tau)=Rf(R\tau)$ where $\tau$ is the interevent time and $R=1/\bar{\tau}$ is the average occurrence rate which depends on magnitude, space scale and different locations. Corral also argued that the optimal fitted function of $f(x)$ is the generalized gamma distribution\cite{Corral2003a}. This scaling function follows a power law for small scales and decays exponentially at large scales. 
Some questions have been raised regarding the universal scaling with region size\cite{Touati2009}. In the context of the epidemic-type after-shock sequence (ETAS) model, the scaling function has been found to depend on the ratio between correlated and independent events\cite{Touati2009,Saichev2006}. In addition, a different study suggested that multiple characteristic time scales, which are controlled by the parameters of the ETAS model, are relevant for the universal scaling behavior of the interevent time distribution in the ETAS model\cite{Bottiglieri2010}.

%The unified distribution of interevent times possibly implies a common mechanism of weak and strong earthquakes at different locations.

%In the epidemic-type after-shock sequence (ETAS) simulation model, every earthquake above a certain magnitude has a probability to trigger other earthquakes based on several basic laws\cite{Ogata1988}. 

%Indeed, the distribution of interevent times can be derived from the dynamics of earthquake i.e. the ETAS model\cite{Saichev2006}. Still, 
The distribution of earthquake events alone does not reflect all the information about the dynamics, and further time series analysis could improve our understanding of the underlying dynamics of earthquakes. For example, Livina et al.\cite{Livina2005} studied the conditional probability of consecutive interevent times and found that these are correlated and not random; i.e., a short interevent time tend to follow a short one and a long interevent tend to follow a long one. Furthermore, detrended fluctuation analysis (DFA) of the interevent interval time series indicated long-range (power law) correlations\cite{Lennartz2008}. In addition, memory has been found bewteen magnitudes of earthquakes\cite{Lennartz2008,Lippiello2008}. Indeed, the conditional probability method and the DFA findings in real data have been recently applied to study and improve the ETAS earthquake model\cite{Fan2018b}.

The studies mentioned above focused mainly on short term memory i.e., between consecutive interevent intervals of earthquakes. However, it is of much interest to test the possibility of long-term memory between interevent intervals. We suggest here to do it by considering the ``lagged'' conditional probabilities. Here, we therefore consider not only the dependence of an interevent interval on the previous one (as done in the past\cite{Livina2005}) but also the conditional probability of an interevent interval depending on a previous (lagged) $k$-th interevent. As we show below, this method extends our understanding regarding the decay of memory with the lag $k$. In addition, we study, for the first time, the memory i.e., the lagged conditional probability, for the interevent distances series.
%The distribution of the interevent distances does not obey the Omori law but depends on the rupture size\cite{Wells1994}.  
%the distribution of the interevent distances is very different from the interevent times one and thus it is important to understand what factors lead to these differences. 
%Although the distribution of the interevent distances is very different from the interevent times, 
The decay with lag $k$ of both interevent times and distances reflect the long term/range memory aspect of the dynamics of earthquakes. Moreover, it is important, as done here, to compare the memory in the interevent times and distances found in real earthquake catalogs to the corresponding memory in a frequently used earthquake forecasting model.

\section*{Memory in the real seismic catalogs}	

\begin{figure}
\centering
\includegraphics[scale=0.8]{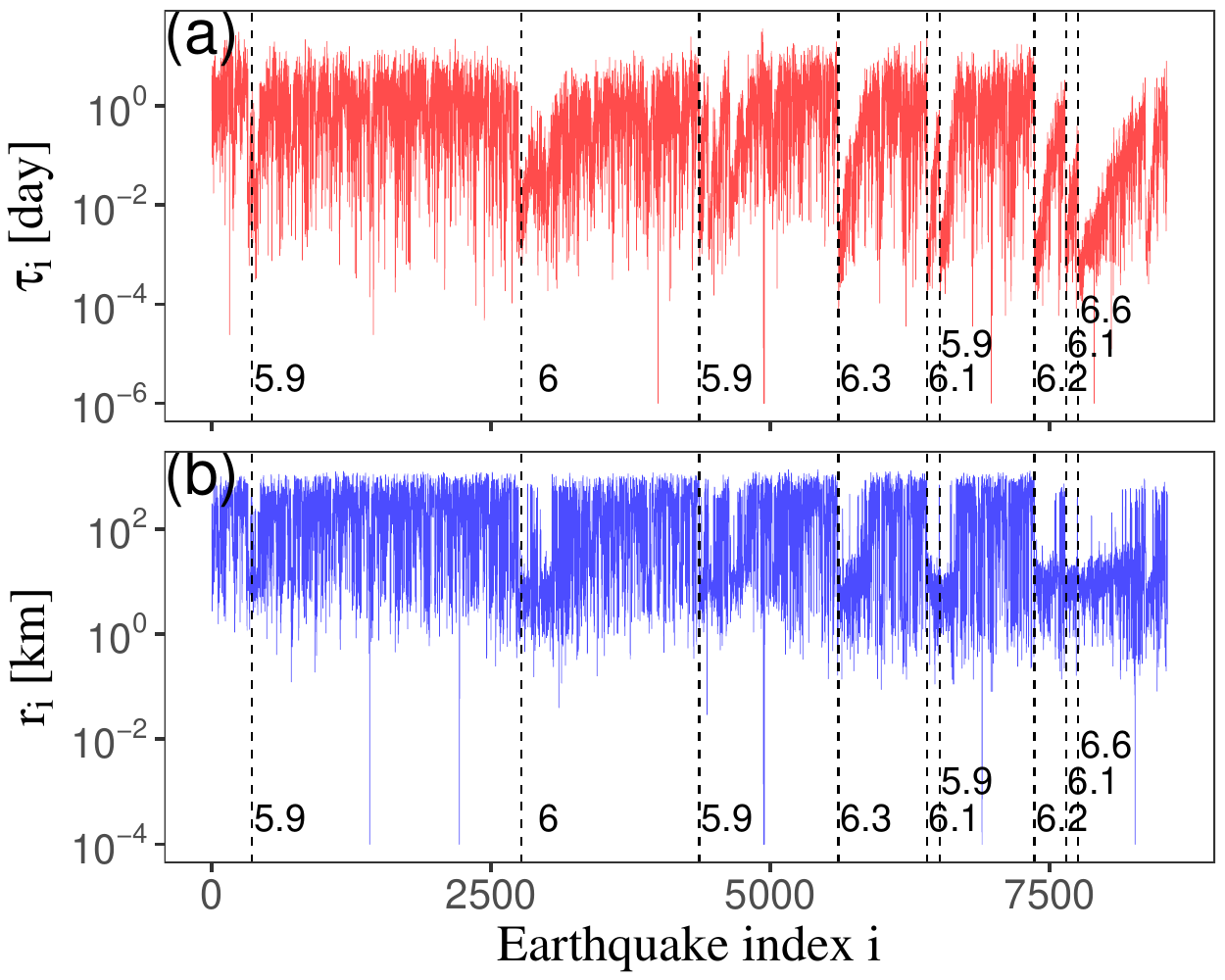}% Here is how to import EPS art
\caption{\label{fig1} Time series (1981--2017) of (a) interevent times and (b) their corresponding distances for the whole the Italian catalog for a magnitude threshold $3.0$. The vertical dashed lines indicate earthquake events with magnitudes above $5.8$. Note the simultaneous decrease in both interevents time and distance appear just after these events.}
\end{figure}

\begin{figure}
\centering
\includegraphics[scale=0.7]{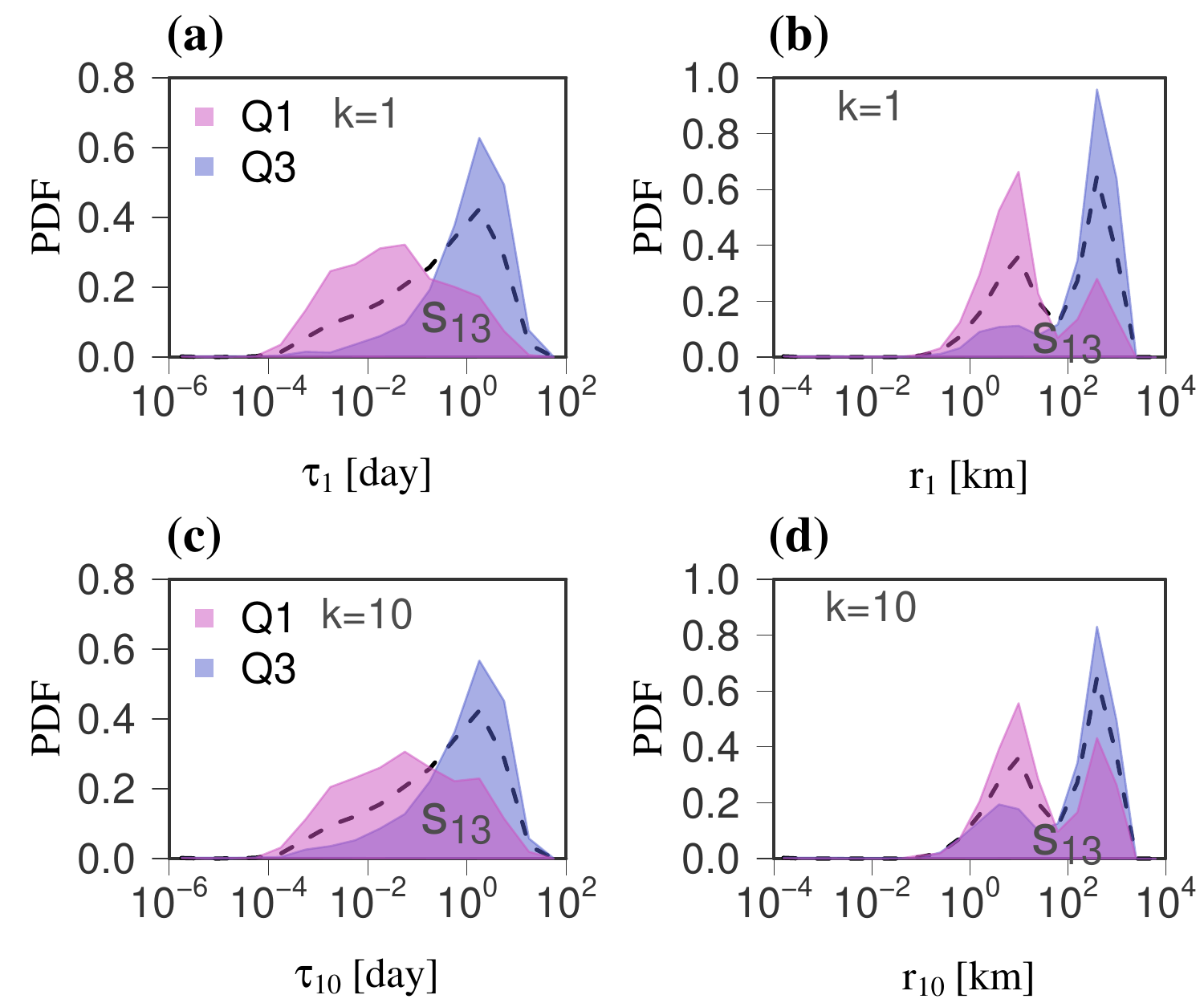}% Here is how to import EPS art
\caption{\label{fig2}(Color online) Conditional PDF of (a) the interevent times $\tau_1$ and (b) the interevent distances $r_1$ (for magnitude threshold $M_0=3.0$) for all Italy; the common area between the conditional PDF of the first and third quantiles is $s_{13}=0.43$ and $s_{13}=0.42$ for times and distances respectively. Note that no memory corresponds to $s_{13}=1$ and full memory to $s_{13}=0$. (c),(d) Same as (a),(b) but for lag $k=10$ where the common area is now $s_{13}=0.57$ and $s_{13}=0.67$ for times and distances respectively. Larger common area indicates less correlations. The black dashed curves indicate the PDFs for all $\tau$ ($r$). The PDFs are normalized in a logarithmic scale.
  % (i.e., the normal normalized PDF $\rho_0(\tau)=\rho(\tau)\frac{d\log_{10}(\tau)}{d\tau}$).
}
\end{figure}

We start by analyzing the seismic catalog of Italy (Methods). An interevent interval time, $\tau_i$, is defined as the time interval between two consecutive earthquake events, $\tau_i=t_{i+1}-t_i$ (in days), above a certain magnitude threshold. Following the Gutenberg-Richter law, the mean interevent time increases with the threshold magnitude. Similarly, we define an interevent distance, $r_i$, as the distance (in kms) between the locations of events $i+1$ and $i$ above a certain magnitude threshold. Figure \ref{fig1} depicts the time series of interevent times and their corresponding distances for all Italy. As seen, after the occurrence of a large earthquake, the interevent times decrease rapidly and then slowly increase, in agreement with the Omori law mentioned above (Fig. \ref{fig1}(a)). Similarly, the interevent distances (Fig. \ref{fig1}(b)) also typically fast decrease after a large earthquake but the following gradual increase observed in the interevent times (see Fig. \ref{fig1}(a)) is less apparent here. We find that similar behavior occur for two specific smaller areas in Italy (see Supplementary Figure 3) and also for Japan and California (Supplementary Figure 4).        
%This difference can be clearly seen from the distribution of interevent times [Fig. S1(a)] and interevent distances [Fig. S1(b)] where the two separated peaks correspond to the short and long distances; we did not observe two separate peaks for the distribution of interevent times.
Supplementary Figure 5 depicts a scatter plot of interevent times vs. interevent distances where two well separated blobs can be observed; see also Figure \ref{fig2}(b). These two groups can be attributed to aftershocks (left blob, short distances) and main-shocks (right blob, long distances); see also\cite{Zaliapin2008,Batac2014}. Supplementary Figure 5 shows that the interevent times and distances exhibit some dependence when considering different blobs (as shorter distances have shorter interevent time) but seem almost uncorrelated within each of the blobs.
%such that analysis of the two may result in new knowledge regarding earthquakes.
 %to some degree, with each other; this is consistent with Fig. \ref{fig1} where it shown that both interevent times and distances become short after strong earthquakes. This clustering underlies the memory in earthquakes\cite{Livina2005}.
 
To study and quantify possible long-term memory in the interevent times and distances of earthquakes, we introduce a lagged conditional PDF method (see Methods). Figure \ref{fig2}(a) shows the lagged $k=1$ conditional PDF, $\rho(\tau_1|\tau_0)$, for the first and third quantiles, $Q1$ and $Q3$. Both $\rho(\tau_1|\tau_0)$ for Q1 and Q3 are substantially different from the overall PDF $\rho(\tau)$ (indicated by the dashed line). These results are consistent with the significant short term memory of the nearest time intervals reported in previous studies\cite{Livina2005,Fan2018b}. In addition, we find significant memory for the interevent distances function $\rho(r_1|r_0)$ as shown in Figure \ref{fig2}(b). Moreover and interestingly, we find significant long-term and long-range memory (correlations) even for large lags, e.g., for $k=10$ (Figure \ref{fig2}(c) and (d)) and for $k=50$ and $k=100$ (Supplementary Figure 6). This type long term memory has not been detected before. The significance of this memory can be verified by comparing the conditional PDF to those of the randomly shuffled time series. Since the randomly shuffled time series contain no memory, $\rho(\tau_k|\tau_0)$ and $\rho(r_k|r_0)$ should be identical to unconditional overall PDF of interevent times and distances, $\rho(\tau)$ or $\rho(r)$, as indeed shown in Supplementary Figure 7. 

To quantify the level of memory expressed by the conditional PDF, we suggest as a measure the common area between the conditional PDFs of the smallest and largest quantiles $Q1$ and $Q3$. When there is no memory the common area should be equal to one while when the PDFs of the two quantiles are completely separated, the common area is expected to be zero. The common area is marked in Figure \ref{fig2} as $s_{13}$. Therefore, we define the level of memory to be $S(\tau_k|\tau_0)\equiv1-s_{13}$ in a range between $0$ to $1$ where large $S(\tau_k|\tau_0)$ indicates strong memory. Similarly, $S(r_k|r_0)$ represents the level of memory for the interevent distances.

%Bak et al\cite{Bak2002} and Corral\cite{Corral2003} found that the mean interevent time is $\bar{\tau} \sim L^{-d_{f}}10^{bM_0}$ where $L$ is the spatial grid size $d_{f}\approx 1.6$ is the 2d fractal dimension of the location of epicenters projected onto the surface of the Earth, $b\approx 1$ is the $b$ value in the Gutenberg-Richter law.% This relation indicates that the interevent times depend on the selections of the grid size and magnitude threshold. Consequently, to understand the dependence of the memory on the grid size and magnitude threshold, Following this relation 

Next we divide the entire Italy into a grid of boxes of edge size $L$ and construct the interevent times (distances) of the events within each box. Then, as described above, we obtain the memory measure $S(\tau_k|\tau_0)$ for all the boxes (grid points) of size $L$. The conditional PDF for the small size $L=3.5^\circ$ is depicted in Supplementary Figure 8. Figure \ref{fig3}(a) shows $S(\tau_k|\tau_0)$ for different grid sizes $L$ (different colors) and magnitude thresholds $M_0$ (different symbols). The memory in the small grid size (red) is stronger than the memory of large grid size (green). The weaker memory for the larger grid size is due to the mixture of the weakly correlated events from remote locations with the nearby highly correlated events. The corresponding memory measure $S(r_k|r_0)$ for the interevent distances exhibits weaker dependence on the grid size (Fig. \ref{fig3}(b)).
% The PDF for large quantile $Q3$ is limited by the size $L$ and as the grid size increases, the PDF of $Q3$ shifts to right and the common area is reduced (Fig \ref{fig2}(b)) indicating stronger memory. [YONGWEN, THE LAST SENTENCE DOES NOT SEEM TO FOLLOW THE RESULTS. I SUGGEST TO EXCLUDE THIS SENTENCE.]
Note that the memory measure, $S$, has two decay rates---it decays slowly for small $k$ and faster for large $k$, both for the interevent times and interevent distances. Moreover, the crossover point for both is nearly the same.  

The crossover from slow decay rate to fast decay rate occurs at smaller $k$ for higher earthquake magnitude threshold (Figs. \ref{fig3}(a) and (b)). This is since the interevent lagged interval of higher magnitude threshold is longer and corresponds to more interevent intervals of smaller magnitude threshold (see Supplementary Fig. 1). Since the frequency of earthquakes decreases exponentially with magnitude (Gutenberg-Richter law), the interevent times grow exponentially with magnitude. Indeed, if we rescale $k$ as $k 10^{bM_{0}}$ (where $b=1$ as for the Gutenberg-Richter law), and multiply the memory measure, $S$, by $L^{d_{u}}/10^{aM_{0}}$ we obtain a single scaling function (Figures \ref{fig3}(c) and (d)).
% In addition, as explained below, the memory measure, $S$, is multiplied by $L^{d_{u}}/10^{aM_{0}}$.
%Indeed, Figs. \ref{fig3}(c) and (d) show the rescaled memory measure $S$ versus rescaled lag $k$; it is seen that all the cases collapse well into the following single function,
Thus, the rescaled memory measure, $F(x)$, is
\begin{equation}\label{eq1}
 F(k10^{bM_{0}})=S(k)L^{d_{u}}/10^{aM_{0}},
% S(k)=L^{-d_{u}}10^{aM_{0}}F(k10^{bM_{0}}),
\end{equation}
where $d_u=0.14$ and $a=0.09$ for the interevent times and $d_u=-0.08$ and $a=0.24$ for the interevent distances. Thus, we are able to rescale the memory measure of different grid sizes and different magnitude thresholds into a single function (memory measure). The scaling parameters were obtained by minimizing the average of the standard variation in all bins in Figs. \ref{fig3}(a) and (b). The average of the standard variation for different parameter values is shown in Supplementary Figure 9.  

It is apparent that the memory measure decreases for larger grid size $L$ and $d_u>0$ represents this decrease for the interevent times. For the interevent distances we find the opposite where $d_u<0$, indicating that memory measure actually increases with the grid size. 
The PDF for large quantile $Q3$ is limited by the size $L$ and as the grid size increases, the PDF of $Q3$ shifts to right and the common area is reduced (Fig \ref{fig2}(b)) indicating stronger memory. 
% is affected by the finite-size effect.
Note that the proposed scaling with area size is not good for very small grid size $L<1^{\circ}$, a size which corresponds to the distance separating the two peaks seen in Fig. \ref{fig2}b. Grid sizes smaller than $\sim 1^{\circ}$ seem to be shorter than typical rapture size of an earthquake, thus breaking the scaling. 
%%Smaller grid size probably divides a aftershock sequence into two separated sequences leading to a difference with the large $L$. 
%When the area size is larger than the spatial scale of aftershock sequence, the scaling behavior could be attributed to that the variation of memory with area size is only affected by independent events.             
We find a positive $a$ indicating that a larger magnitude threshold $M_{0}$ tends to have a stronger memory after $k$ is rescaled. Note that the parameter $a$ of interevent distances is larger than the corresponding $a$ parameter of interevent times.

The scaling functions shown in Figs. \ref{fig3}(c) and (d) indicate a crossover between two distinct power law relations. The scaling function is $F(x) \sim x^{-\gamma_1}$ for $x=k10^{bM_{0}}$ where $x$ is in the range of $[10^0,10^{5}]$ and $\gamma_1$ is $0.19$ and $0.21$ for the interevent times and interevent distances respectively. Both scaling functions exhibit a significant crossover at $x_c\approx 10^{5.0}$ and the approximate scaling exponent for large $x$ (i.e., in range of $[10^5,10^{5.5}]$) is $\gamma_2=0.88$ and $1.11$ for interevent times and interevent distances respectively. It is clear that $\gamma_1\ll\gamma_2$ such that the decay for small scales is much slower than the decay for larger scales. Note, that due to the short range of large $x$ $[10^5,10^{5.5}]$ this range decay could be also fitted to an exponential decay.

Supplementary Figure 10 shows the relation between the average (and rescaled average) time differences between two earthquakes and their lag $k$ (and rescaled lag). It indicates that the crossover for the smaller grid size $L$ corresponds to a longer time. For the entire Italy, the crossover time, $t_{c}$ is around 130 days and it is the same for different magnitude thresholds.
% ; the crossover time indicates that there is a time scale that characterizes earthquake activity.

To verify the generality of our scaling, we performed the same scaling analysis (using Eq. (\ref{eq1})) for Japan and California earthquake catalogs and obtained also good scaling. The scaling functions exhibit a similar crossover to that discussed above (Supplementary Figs. 11, 12). The scaling parameters and exponents are slightly different for the different catalogs and are summarized in Table \ref{T1}. The rescaled memory measure $F$ of Japan catalog decays slower (as expressed in the smaller exponents $\gamma_1$ and $\gamma_2$) in comparison to other locations, probably due to the high earthquake rate there. The rescaled lags of the crossover, $x_c$, are also listed in Table \ref{T1}. Note that the distance and time have similar values for the crossover, $x_c$, for all three regions. Due to the high earthquake rate, the crossover time $t_c$ is also shortest for the Japanese earthquake catalog (see Table \ref{T1}).      

\begin{figure}
\centering
\includegraphics[scale=0.5]{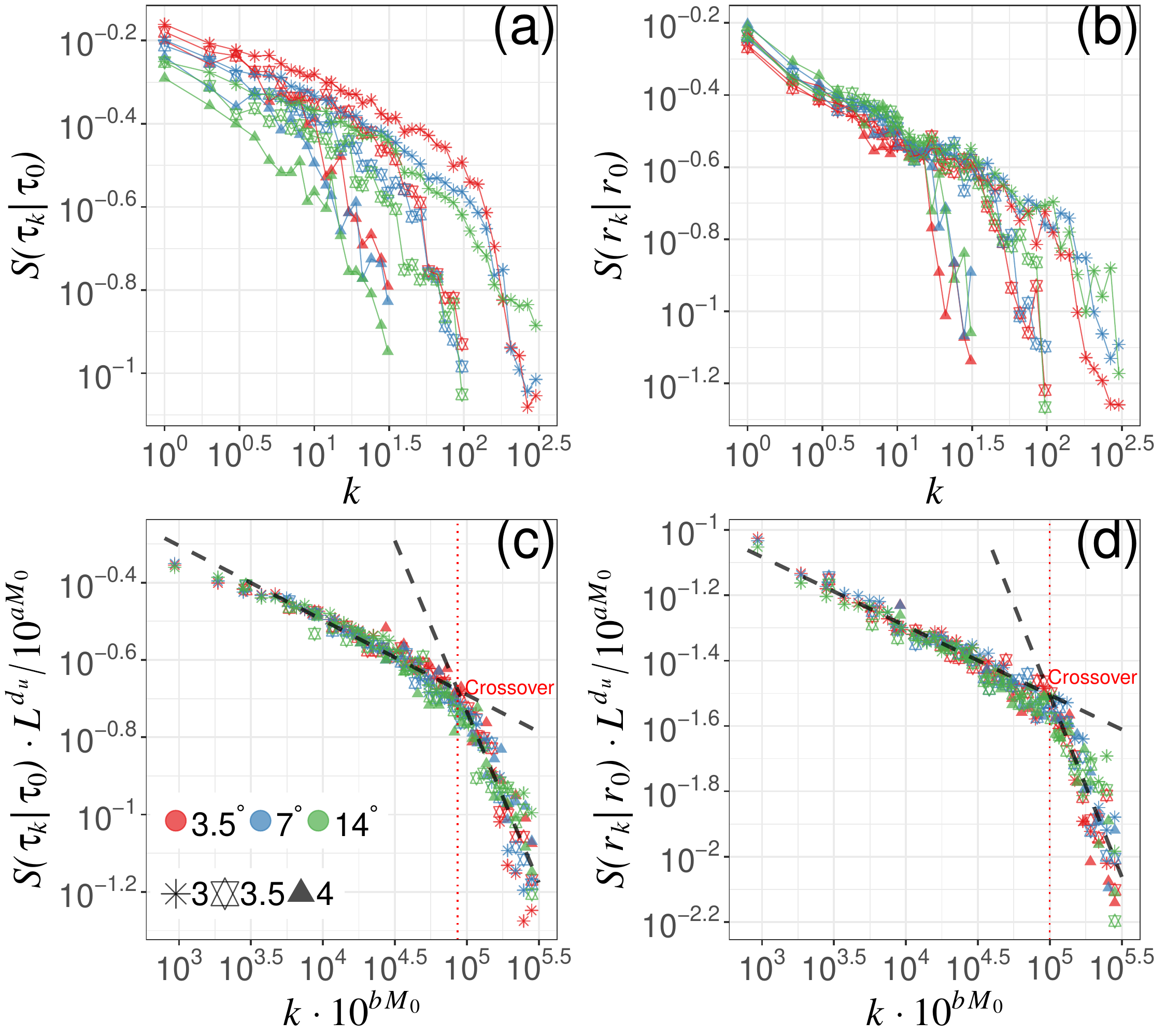}% Here is how to import EPS art
\caption{\label{fig3}(Color online) The memory measure (a) $S(\tau_{k}|\tau_{0})$ and (b) $S(r_{k}|r_{0})$ as a function of the lag index $k$ for interevent times and distances respectively. Colors represent different grid sizes, $L$. Shapes of symbols represent different magnitude thresholds, $M_0$. The grid size of $L=14^\circ$ covers the whole region of the Italian catalog. Rescaled memory measure (c) for interevent times and (d) distances. Black dashed lines are fitted power-law curves. Note that the narrow large $k$ regimes could be fitted also by exponential decay. The vertical red dotted lines indicate the location of the crossover.}
\end{figure}

\begin{table}%The best place to locate the table environment is directly after its first reference in text
\caption{\label{T1}%
Estimated parameters, $a$, $d_u$, power law exponents of the scaling function, $\gamma_1$,$\gamma_2$, and the crossover points, $x_c$, $t_c$ for the interevent times and distances for Italy, Japan and California earthquake catalogs.}
\centering
\begin{tabular}{ccccc}
\toprule
\textrm{}&
\textrm{Parameters}&
\textrm{Italy}&
\multicolumn{1}{c}{\textrm{Japan}}&
\textrm{California}\\
\midrule
 & $a$ & 0.09 & 0.07 & 0.1\\
 & $d_u$ & 0.14 & 0.05 & 0.07\\
 & $\gamma_1$ &0.19 &0.09 &0.2\\
Time & $\gamma_2$ &0.88 &0.28 &0.65\\
 & $\gamma_2/\gamma_1$ &4.63 &3.11 & 3.25\\
 & $log_{10}(x_c)$ & 4.91 & 4.77 & 4.98\\
  & $t_c$ & 126 days & 19 days & 88 days\\
\midrule
 & $a$ & 0.24 & 0.2 & 0.2\\
 & $d_u$ & -0.08 & -0.07 & -0.14\\
 & $\gamma_1$ &0.21 &0.23 &0.35\\
 Distance & $\gamma_2$ &1.11 &0.71 &0.75\\
 & $\gamma_2/\gamma_1$ &5.29 & 3.09& 2.14\\
 & $log_{10}(x_c)$ & 4.97 & 4.74 & 4.97\\
 & $t_c$ & 144 days & 18 days & 86 days\\
 \bottomrule
\end{tabular}
\end{table}

\section*{Memory in the ETAS model}	

\begin{figure}
\centering
\includegraphics[scale=0.5]{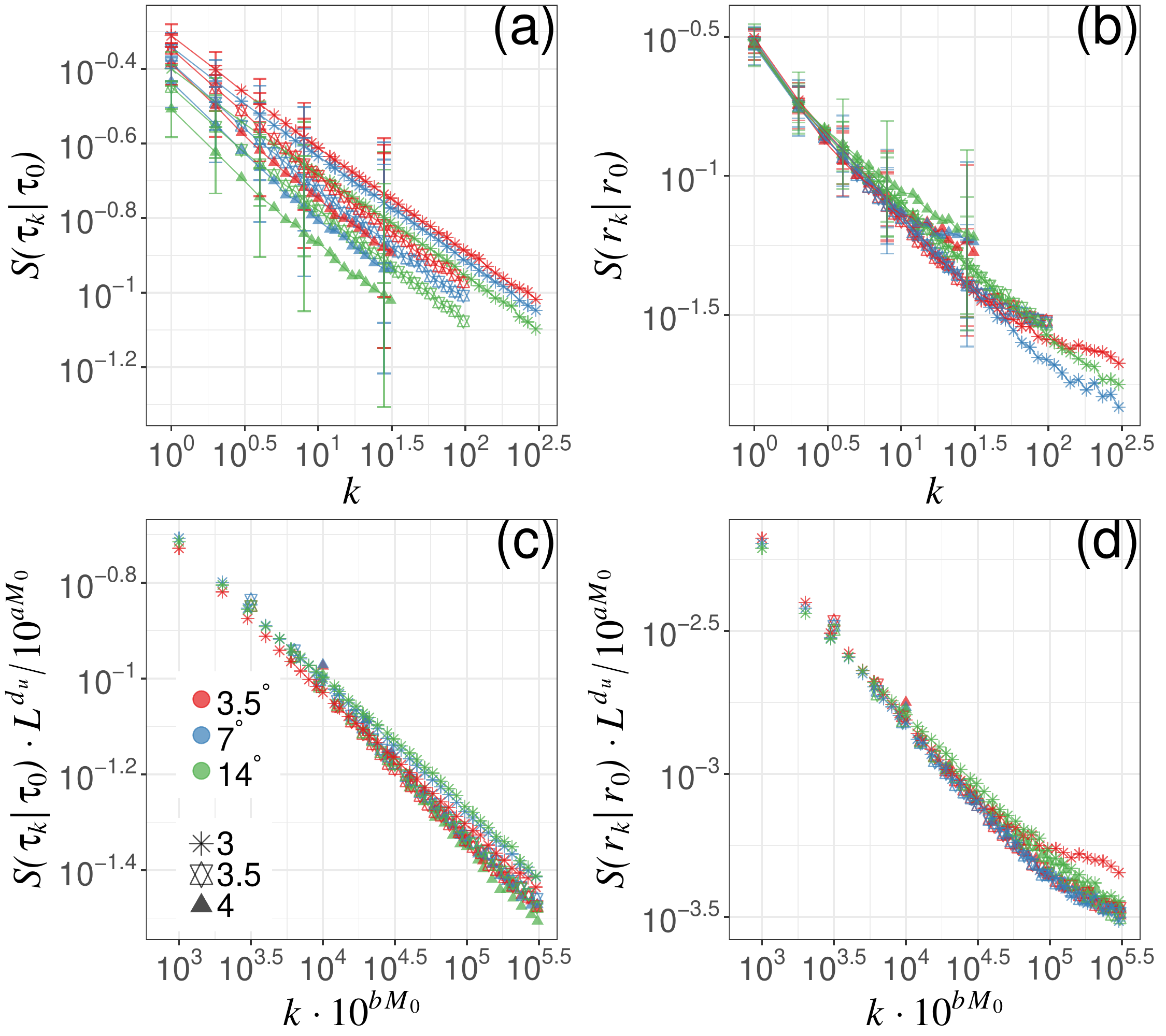}% Here is how to import EPS art
\caption{\label{fig4}(Color online) Memory measure obtained from the output of the ETAS model (a) $S(\tau_{k}|\tau_{0})$ (interevent times) and (b) $S(r_{k}|r_{0})$ (interevent distances) as a function of the lag-index $k$. Colors and shapes are same as in Fig. \ref{fig3}. Rescaled memory measure for (c) interevent times and (d) distances. The model's parameters where estimated for the Italian catalog\cite{Fan2018b, Lombardi2015} and are $\mu=0.2$, $A=6.26$, $p=1.13$, $c=0.07$ and $\alpha=1.5$. The memory measure $S$ is averaged over $50$ independent realizations and each realization includes $10^6$ events. The error bars are the standard deviations.}
\end{figure}

A good earthquake model should be able to reproduce the observed long-term and long-range memory features. Such a model could have the potential to significantly improve the forecasting capability of earthquakes. Thus, we next test the memory in the frequently used earthquake model, the epidemic-type after-shock sequence (ETAS) model (Methods)\cite{Ogata1988, Ogata1998}. The dynamic of the ETAS model is known as a stochastic space-time point process. Every earthquake above a certain magnitude has a probability to trigger other earthquakes based on several basic laws\cite{Ogata1998}. The ETAS model can provide statistically some reliable forecasts of seismicity\cite{Marzocchi2017}. Supplementary Figure 13 shows the conditional PDF of the interevent times and distances and these exhibit much larger overlap in comparison to the memory in the data (Figure \ref{fig2}), indicating weaker memory in the model. Figures \ref{fig4}(a) and (b) shows the memory measure for both interevent times and distances calculated from the ETAS model. As for the real catalogs (Figs. \ref{fig4}(c), (d)), the memory measure of the model satisfies the scaling relation expressed in Eq. (\ref{eq1}). The scaling parameters are $d_u=0.17$, $a=0.17$ for the interevent times and $d_u=-0.04$, $a=0.55$ for the interevent distances. As expected, the grid size scaling lead to $d_u>0$ for the interevent times and $d_u<0$ the interevent distances, similar to the real catalogs. Also, the magnitude threshold scaling parameter $a$ is positive, as for the real catalogs.
% Also, the rescaled memory measure of the model depends on the magnitude threshold as $a>0$.
Yet, the value of $a$ of the interevent distances in the model ($a=0.55$) is 2-3 times larger than the value we obtain for the real catalogs ($a=0.24$, Table \ref{T1}). Thus, the memory measure of the model is significantly smaller than that of the real catalog, even when using the optimal parameters for the ETAS model suggested in\cite{Fan2018b, Lombardi2015}. Moreover, the crossover power-law behavior at $x_c$ that has been observed in the real catalogs (Fig. \ref{fig3}) can not be seen in the model (Fig. \ref{fig4}). We thus conclude that the ETAS model does not reproduce the main memory features found in the present study for real catalogs.

To better understand which parameters controls the power law exponent and to find the optimal model's parameters that reproduce the characteristics of the real catalogs, we perform sensitivity tests by varying the $p$ and $\alpha$ parameters for different choice of $\mu$. We estimate the scaling exponents $\gamma_1$, $\gamma_2$ for different regimes of $k$ (similar to Figs. \ref{fig3}c, d and Table \ref{T1}). The ratio between the high and small $k$ exponents, $\gamma_2/\gamma_1$ quantifies the level of crossover and the double power-law behavior. For real catalogs the ratio $\gamma_2/\gamma_1$ is much larger than 1 and is about 5 for the interevent times and above 2 for the interevent distances.
For the model, the results of the sensitivity tests for interevent times and distances are shown in Supplementary Figures 14 and 15. It is seen that the ETAS model fail to reproduce the large $\gamma_2/\gamma_1$ observed in the real catalog; the largest $\gamma_2/\gamma_1$ is about $1.0$ (see Supplementary Figs. 14, 15).

%The parameter $p$ could control the power law exponent for the interevent times by the Omori law so it does not work for the distance. [IMPROVE THE LAST SENTENCE.]
%However, the power law exponent of the time and distance both significantly respond to the productivity parameter $\alpha$.
\section*{Discussion} 
Although the distributions of interevent times and interevent distances of the ETAS model are similar to the distributions obtained from the real catalogs, the dynamics of earthquakes represented by memory is found here not to be captured by the ETAS model. We find that the memory in the model is weaker and decays faster in the short time scale compared to the real catalogs. For the long term, the memory in the model decays slower than for the real catalogs. The model does not exhibit the clear crossover observed in the real catalogs. The two different power law decays found here in the real catalogs and the clear crossover between them, may imply the existence of large and small productivity rates $\alpha$ corresponding to short and long term memory in the dynamics of earthquakes. We conjecture, based on the above, that an improved ETAS model should include two $\alpha$ values for short and long term triggering. 
%For the short term triggering, a large $\alpha$ is taken as possible and for the long term triggering, a small $\alpha$ is used to replace. [THE LAST SENTENCE IS NOT CLEAR.]

In summary, we propose and analyze, for the first time, a lagged conditional PDF method, and detected significant memory in earthquake catalogs in both interevent times and interevent distances. The memory depends on the magnitude threshold and area size (grid size). We propose a scaling function for which the different conditional probability curves collapse into a single scaling function for different magnitude thresholds and area sizes; the scaling has been observed successfully on earthquake catalogs from Italy, Japan and California. The rescaled memory measure indicates a clear crossover between two power law relations as a function of the rescaled lagged-index $k10^{bM_0}$. We also find that the power-law exponents strongly depend on the ETAS model productivity parameter $\alpha$ both for interevent times and distances. However, the double power law behavior (crossover) found here for the real catalogs is not observed in the ETAS model and the memory in the ETAS model is weaker in comparison to that found here for real data. This finding may imply the need to extend the ETAS model to include two different productivity rates ($\alpha$) for short and long term memory as we plan to perform in a forthcoming study.

\section*{Methods}

\subsection{Data}
The Italian earthquake catalog is provided by the Seismic Hazard Center at Istituto Nazionale di Geofisica e Vulcanologia (INGV, http://terremoti.ingv.it/it/). This catalog spans the time period from 1981 to 2017 and it is complete for earthquake magnitudes above $M_0=3.0$; a catalog is ``complete'' when all events above the specified magnitude are included in the catalog (see Fig. S1). The spatial distribution of earthquakes for the catalog is shown in Figure S2. The Japan catalog is the Japan Unified High-Resolution Relocated Catalog for Earthquakes from 2001 to 2012 (JUICE, http://www.hinet.bosai.go.jp/topics/JUICE/?LANG=en) and the California catalog is provided by the California Integrated Seismic Network for 1970--2018 (CISN, https://earthquake.usgs.gov/data/\\comcat/contributor/ci).

\subsection{The conditional PDF}

The conditional PDF method generalizes the method introduced by\cite{Livina2005} and implemented by\cite{Fan2018b}. First, we sort the all interevent times (distances) in ascending order and then divide the sorted series into three equal quantiles. Thus, the first quantile, $Q1$, contains the smallest $1/3$ interevent times (distances) and the third quantile, $Q3$, contains the largest $1/3$ interevent times (distances). We define the conditional PDF of interevent times and distances as $\rho(\tau_k|\tau_0)$ ($\rho(r_k|r_0)$), where $\tau_0$ ($r_0$) belongs to $Q1$ or $Q3$, and $\tau_k$ ($r_k$) is the lagged $k$-th interevent time that follows $\tau_0$ ($r_0$). Note that earlier studies\cite{Livina2005, Fan2018b} considered only the first lag ($k=1$) and considered only interevent times (but not interevent distances).

\subsection{The ETAS Model}
In the ETAS model, seismic events are assumed to involve a Poisson process where the conditional intensity function $\lambda$ is controlled by a few estimated parameters\cite{Touati2009,Fan2018b} as follows, 
\begin{equation}\label{eq3}
\lambda(t|H_t)= \mu + A\sum_{t_i<t}\exp(\alpha(M_i-M_0))\left(1+\frac{t-t_i}{c}\right)^{-p}\;,
\end{equation}
where $H_t$ is the history process before time $t$, $t_i$ are the times of the past events and $M_{i}$ are their magnitudes. The parameter $\mu$ is the background event rate, $A = K/c^{p}$ is the occurrence rate of earthquakes in the Omori law at zero lag, $c$, $p$ and $K$ are the Omori law parameters, and $\alpha$ is the productivity parameter. The function $\lambda(t|H_t)$ is the probability at time $t$ for the occurrence of an earthquake with a magnitude threshold above $M_0$, given the earthquake history $H_t$. Each event's magnitude is selected independently from the Gutenberg--Richter distribution. The branching ratio $n$ is the average number of events triggered by each event. It can be obtained by integrating $A\exp (\alpha M)\left(1+\frac{t}{c}\right)^{-p}$ over both time and magnitude from $0$ to $\infty$ and is given by $n=\frac{A c}{p-1} \frac{\beta}{\beta-\alpha}$, where $\beta=b\ln10$ ($b=1$ on the Gutenberg-Richter law). To avoid ETAS model's singularities and to have physical stability, one assumed $0<n<1$, such that $p>1$, $\alpha<\beta$ and $n<1$ \cite{Sornette2002}. 

The spatial coordinates $x$ and $y$ of earthquakes in the ETAS model can be obtained independently. the conditional intensity $\lambda(x, y, t|H_t)$ can be integrated using spatial coordinates $x$ and $y$ to get the total conditional intensity for the entire region that is equal to the temporal one.%‐-only ETAS conditional intensity in Eq. (2).
Thus, times and spatial coordinates can be obtained separately\cite{Touati2011a}: (i) times are calculated for the entire region by Eq. (2); (ii) then we choose spatial coordinates in the region for the events. For simplicity, spatial background events' are assumed to be spatially random. Spatial aftershock events' coordinates $x$ and $y$ are obtained using a spatial kernel function\cite{Zhuang2012}:
\begin{equation}\label{eq4}
f\left(x-x_i, y-y_i, M_{i}\right)=\frac{q-1}{\pi D\zeta(M_i)}\left(1+\frac{(x-x_i)^{2}+(y-y_i)^{2}}{D\zeta(M_i)}\right)^{-q},
\end{equation}
\begin{equation}\label{eq4}
\zeta(M_i)=\exp \left[\gamma_m\left(M_{i}-M_{0}\right)\right],
\end{equation}
where $i$ represents the triggering event; $q$, $D$ and $\gamma_m$ are the estimated parameters. In the present study the value of these parameters are $q=1.66$, $D=0.001$ and $\gamma_m=0.64$ for the Italian catalog; these were estimated using the maximum likelihood estimator\cite{Zhuang2012}.

%% Put the bibliography here, most people will use BiBTeX in
%% which case the environment below should be replaced with
%% the \bibliography{} command.
%\bibliographystyle{natbib}
%\bibliography{earthquake}

%\begin{thebibliography}{1}
%\bibitem{dummy} Articles are restricted to 50 references, Letters
%to 30.
%\bibitem{dummyb} No compound references -- only one source per
%reference.
%\end{thebibliography}

%% Here is the endmatter stuff: Supplementary Info, etc.
%% Use \item's to separate, default label is "Acknowledgements"

\begin{addendum}
 \item We thank Xiaosong Chen for helpful discussions. We thank the Italian Ministry of foreign affairs and international cooperation, and the Israeli Ministry of science, technology, and space; the Israel Science Foundation, ONR, Japan Science Foundation, BSF-NSF, ARO, the EU H2020 project RISE, and DTRA (Grant no. HDTRA-1-10-1-0014) for financial support. 
 \item[Competing Interests] The authors declare that they have no
competing financial interests.
 \item[Correspondence] Correspondence and requests for materials
should be addressed to Yongwen Zhang (email: zhangyongwen77@gmail.com).
\end{addendum}

%%
%% TABLES
%%
%% If there are any tables, put them here.
%%

\end{document}